\documentclass[10pt,letterpaper]{article}
\usepackage{opex3}

\usepackage{amssymb}
\usepackage{amsmath}
\usepackage{epsfig}

\def\eq#1{Eq.~(\ref{#1})}
\def\fig#1{Fig.~\ref{#1}}

\begin{document}

\title{Reconfigurable quantum metamaterials}

\author{James Q. Quach$^1$, Chun-Hsu Su$^1$, Andrew M. Martin$^2$, Andrew D. Greentree$^2$ and Lloyd C. L. Hollenberg$^1$}

\address{$^1$ Centre for Quantum Computer Technology, School of Physics, The University of Melbourne, Victoria 3010, Australia}
\address{$^2$ School of Physics, The University of Melbourne, Victoria 3010, Australia}
\email{jamesq@unimelb.edu.au}

\begin{abstract*}
By coupling controllable quantum systems into larger structures we introduce the concept of a quantum metamaterial. Conventional metamaterials represent one of the most important frontiers in optical design, with applications in diverse fields ranging from medicine to aerospace. Up until now however, metamaterials have themselves been classical structures and interact only with the classical properties of light. Here we describe a class of dynamic metamaterials, based on the quantum properties of coupled atom-cavity arrays, which are intrinsically lossless, reconfigurable, and operate fundamentally at the quantum level. We show how this new class of metamaterial could be used to create a reconfigurable quantum superlens possessing a negative index gradient for single photon imaging. With the inherent features of quantum superposition and entanglement of metamaterial properties, this new class of dynamic quantum metamaterial, opens a new vista for quantum science and technology.
\end{abstract*}




\section{Introduction}

By offering material properties beyond that which occurs in nature, artificially engineered metamaterials are of intense interest. Typically fabricated with periodic features spaced closer than the operating wavelength, the system acts as a homogenous material. The earliest introductions were in the area of magnetic resonance imaging where conducting elements were used to produce \emph{artificial magnetism} ~\cite{pendry99,wiltshire01,wiltshire03a,wiltshire03b,yen04}. Negative index materials (NIMs) with simultaneous negative permittivity and permeability have also been engineered. A remarkable property of such negative index materials (NIMs) is the possibility of negative refraction ~\cite{veselago68}. Negative refraction arises due to the ability to design an interface between concave and convex surfaces in the material bandstructure.  This interface leads to a novel dispersion relation not found in naturally occurring materials, namely all-angle negative refraction~\cite{paul09}. Simply put, conservation of energy forces a reversal of the  wave-vector in one dimension, whilst preserving the wave-vector in the other dimension.  Typically, the appropriate bandstructure engineering requires control over the electromagnetic properties on scales significantly smaller than the wavelength of the radiation, although any system that realizes the appropriate bandstructure engineering will exhibit negative refraction. NIMs with  negative refraction have been demonstrated in the microwave regime in structures that consists of interlocking metal strips and conducting split-ring resonators ~\cite{smith00,shelby01a,houck03,parazzoli03}. By applying transformation optics, metamaterials become a rich platform for the control of electromagnetic waves. A striking consequence of such control has been the realization of an invisibility cloak~\cite{schurig06}. Apart from electromagnetic metamaterials, acoustic~\cite{Liu00,Guenneau07} and seismic~\cite{Guenneau09} metamaterials are also areas of intense research.

\begin{figure}[tb]
	\centering
	\includegraphics[width=0.8\columnwidth]{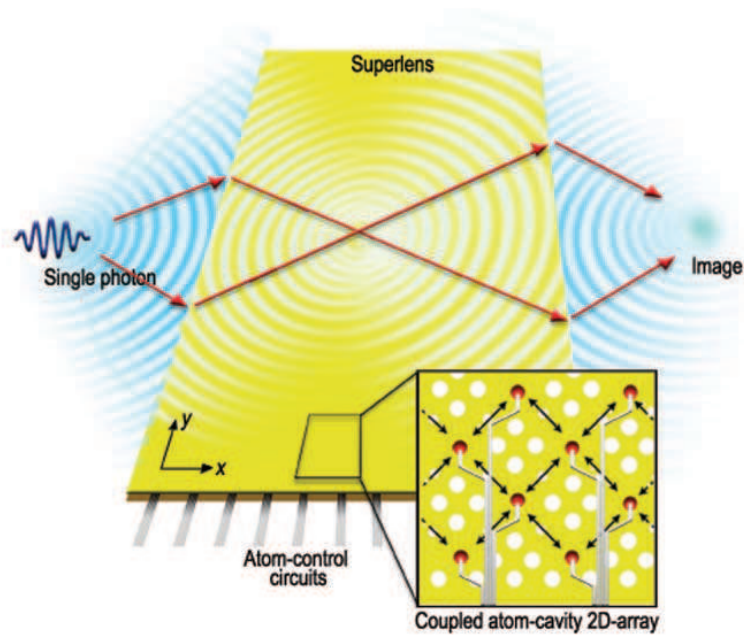}
	\caption{A reconfigurable quantum superlens built from a cavity-array metamaterial. By allowing all-angle negative refraction and evanescent wave enhancement, the superlens surpasses the diffraction limitation of conventional lenses. Inset: Electrostatic tuning of the intracavity atoms (solid circles) in the cavity lattice provides dynamic control over the light guiding and resonance properties of the lens.}
	\label{fig:schematic}
\end{figure}

Here we introduce a new class of metamaterials distinguished from conventional metamaterials in that it operates in the quantum regime and is easily reconfigurable. The medium, a cavity array metamaterial (CAM), comprises of a network of coupled atom-optical cavities. Under rotating-wave approximation and in the tight-binding regime, the medium can be treated with the Jaynes-Cummings Hubbard (JCH) model~\cite{hartmann06,Greentree06a,angelakis07b,zhou08a,makin08,schmidt09,pippan09,quach09}. Atom-optical cavities have already been shown to demonstrate such quantum effects as quantum collapse and revival~\cite{rempe87}, Rabi oscillations~\cite{brune96}, photon blockade~\cite{birnbaum05}, and electromagnetically induced transparency~\cite{mucke10}. The coupling of these cavities have theoretically been shown to exhibit quantum phase transitions~\cite{Greentree06a, quach09} and quantum chaos~\cite{hayward10}. They have also been proposed to act as Q-swtiches~\cite{su09}, single photon switches~\cite{zhou08b}, many-body~\cite{tomadin10} and semiconductor simulators~\cite{quach09}. Importantly, by electrical control of the atomic resonances, the metamaterial properties can be dynamically varied, a significant development compared to the more usual static implementations. The term quantum metamaterial was first introduced in the context of superconducting charge qubits inside a superconducting resonator~\cite{rakhmanov08}. Here we extend the concept of a quantum metamaterial by providing an alternative quantum platform based on cavity quantum-electrodynamics.  This platform goes beyond classical electromagnetic properties and allows the exploitation of a quantum plurality of metamaterial properties. 

As an illustration we present the design for a reconfigurable \emph{quantum superlens} based on the CAM. We envisage a configuration depicted schematically in Fig.~\ref{fig:schematic} where the JCH system is manipulated to produce a perfect image using single photons. The aim here is not build a better superlens, but to use the superlens as an casestudy of how CAMs can be used to exhibit metamaterial properties such as negative refraction. However as with photonic crystal (PhC) implementations, CAMs have the advantage of being ideally lossless and do not require the operating wavelength to be larger than the constitutent element spacing. Further, because the transition energy of each atom in the system can be individually controlled via a Stark shifting control voltage, CAMs have the distinct advantage of being highly tunable and reconfigurable: features not possible using conventional designs.

To form a perfect image requires the lossless convergence of the propagating and evanescent light components. Conventional lenses only focus the propagating fields, and so the resolution of the image is fundamentally limited to features greater than the optical wavelength. Subwavelength features are carried by the high spatial-frequency components encoded by the evanescent fields. The loss of the evanescent components lead to the \emph{diffraction limit}. Near-field scanning optical microscopy overcomes this problem by scanning a probe in close proximity to the object, but this is often undesirable for applications such as optical lithography and sensing. It has been proposed that a lens built from NIMs can produce \emph{perfect} far field imaging, exhibiting all-angle negative refraction (AANR) and evanescent wave enhancement (EWE)~\cite{pendry00}. Because of their ability to overcome the diffraction limit, and the lack of optical axis and curved surfaces that AANR affords, such a lens is termed \emph{superlens}~\cite{pendry00,zhang08}.

Negative indexing however is not a prerequisite for superlensing. A different class of metamaterial from that of NIM, is formed by photonic crystals (PhCs) which uses Bragg scattering. PhCs have also been shown to exhibit AANR and EWE~\cite{luo02,luo03a,luo03b}, and have the advantage of low loss. The disadvantage of PhCs as a superlens is that not all the evanescent components can be uniformly amplified. In relation to the PhC as a metamaterial, it is of note that since the size and periodicity of the scattering elements in PhCs are on the order of the operating wavelength, the medium cannot be considered as homogeneous, which is a necessary condition to identify a meaningful permeability and permittivity. 

The theoretical developments of superlenses have also been matched by experimental efforts. Superlensing has been demonstrated with microwaves in both NIMs~\cite{grbic04} and PhCs~\cite{cubukcu03a,cubukcu03b}, as well as other platforms such as silver films in visible light~\cite{liu03}. More recently, near-field microscopy using a SiC-based superlens at mid-infrared frequency has successfully imaged features smaller than the illumination wavelength~\cite{taubner06}.

We will for the first time discuss cavity arrays as a metamaterial. The inter-cavity hopping mechanism of the CAM are markedly distinguished from NIM and Bragg scattering, which have been the hallmark of metamaterials until now, and therefore represent another class of metamaterial. In particular we will discuss the JCH system as a CAM and investigate it as a medium for a quantum-based reconfigurable superlensing device. The tunability of the system enables post fabrication dynamical control over the focal point of the lens, and by scanning over resonances, importantly allows uniform enhancement of evanescent modes, which is not possible with passive PhC-based devices. We further show a gradient negative index lens which minimizes reflection from the device.

\section{Jaynes-Cummings-Hubbard Hamiltonian}

We consider a uniform two dimensional (2D) periodic array of coupled optical cavities, embedded with single two-level atomic systems.  Our treatment is implementation independent and could be realized in any coupled-cavity array system that realizes a two-dimensional square lattice topology.  However it is useful to consider a concrete implementation.  We specifically consider a coupled-cavity array realized in a two-dimensional photonic crystal membrane.  In such a structure the thickness of the membrane effects confinement in the third dimension, whilst the photonic crystal structure defines the cavity array.  Such a structure has been considered previously in the context of photonic quantum emulators~\cite{Greentree06a,makin08,quach09}, and one-dimensional coupled cavity arrays have been fabricated~\cite{notomi08}. The cavity array is a series of quantum oscillators coupled through the overlapping of the photonic modes of adjacent cavities, such that it leads to a tight-binding Hubbard-like model. Each cavity couples to its atom via the Jaynes-Cummings (JC) interaction. In terms of the atomic (photonic) raising and lowering operators $\sigma^+_r, \sigma^-_r$ ($a_r^\dagger, a_r$) at site $r$, the total JCH Hamiltonian reads ($\hbar=1$),.
\begin{equation}
	\mathcal{H} = \sum_r{ \epsilon\sigma^+_r\sigma^-_r + \omega a_r^\dagger a_r + \beta(\sigma_r^\dagger a_r + a_r^\dagger \sigma_r^-)} - \sum_{\langle r, s \rangle}{\kappa a_r^\dagger a_s},
\label{eq:JCH}
\end{equation}
where $\sum_{\langle r, s \rangle}$ is the sum over all nearest-neighbor cavities, $\kappa > 0$ is the hopping frequency (requiring that the ground eigenstate to be symmetric and the first excited eigenstate to be anti-symmetric also means that $\kappa > 0$), $\epsilon$ is the atomic transition energy, $\omega$ is the cavity resonance frequency, $\beta$ is the single-photon Rabi frequency, and the rotating wave approximation is assumed. The onsite terms can be diagonalized in a basis of mixed photonic and atomic excitations called dressed states or polaritons, $|\pm,n\rangle_r = \sin\Theta_n|g,n\rangle_r + \cos\Theta_n|e,n-1\rangle_r$, with energy $E_n^{\pm} = n\omega - \Delta/2 \pm \sqrt{n\beta^2 + (\Delta/2)^2}$, and mixing angle $\Theta_n = \frac{1}{2}\arctan[-2\sqrt{n}\beta/\Delta]$, where $\Delta \equiv \omega-\epsilon$. 

\begin{figure}[htbp]
	\centering
	\includegraphics[width=0.8\columnwidth]{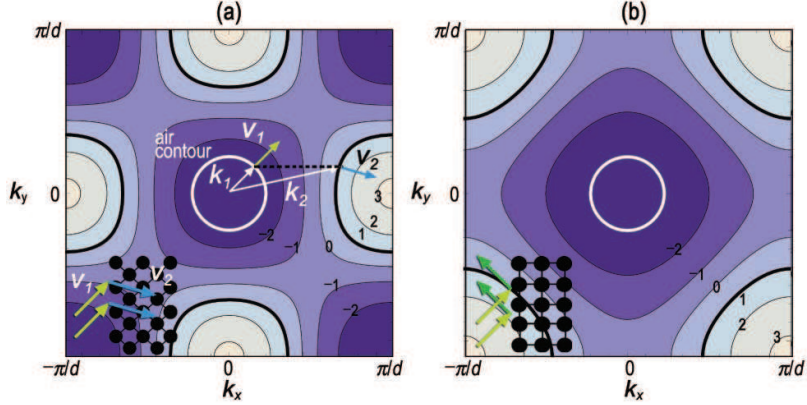}
	\caption{Energy band structure of a JCH lattice with air-lattice isoenergy contours in the first Brillouin zone. Bold lines are the isoenergy contours for an operating frequency in free space (white) and in the lattice (black). White arrows denote the wavevectors $\vec{k}$ parallel to the phase velocity and colored arrows the group velocity $\vec{v}$. At the interface, the $k_y$ component is conserved. (a): In the rotated lattice there is all-angle negative refraction (yellow arrow denotes the incident group velocity, blue arrow the refraction group velocity). (b): In the unrotated lattice, there are no operating frequency propagating modes where $k_y$ is conserved at the interface, and the photon is reflected (green arrow).}
\label{fig:airlatt}
\end{figure}

Using a Bloch state analysis (see \textsc{appendix}) the band structure in the one-excitation manifold is given by,
\begin{equation}
	E^\pm = \frac{1}{2}(\omega + \epsilon - K) \pm \frac{1}{2}\sqrt{(\Delta - K)^2 + 4\beta^2}~.
\label{eq:energy_spectrum}
\end{equation}
For the rotated lattice (depicted in inset of \fig{fig:airlatt}(a)), $K \equiv 4\kappa\cos(k_x d)\cos(k_y d)$ and for the unrotated lattice (inset of \fig{fig:airlatt}(b)) $K \equiv 2\kappa[\cos(k_x d)+\cos(k_y d)]$, where $d\equiv|\vec{d}_s - \vec{d}_r|/\sqrt{2}$. Equation (\ref{eq:energy_spectrum}) is an exact solution for a periodic lattice. In our numerical simulations we will verify that boundary effects are negligible.  Note that \eq{eq:JCH} describes the underlying connectivity of the lattice, and not explicitly its geometry.  It is possible for geometrically distinct configurations to have the same Hamiltonian, and in the reciprocal lattice the same Bloch analysis would hold, resulting in equivalent lensing properties. However, the orientation of the lattice relative to the interface is of importance for superlensing, as is explained below.

In the following sections we will demonstrate how traditional metamaterial techniques can be applied to the JCH Hamiltonian, to show that a suitably designed CAM 
 can exhibit the hallmarks of a superlens, namely all-angle negative refraction and evanescent wave ehancement.

\section{All-Angle Negative refraction}
A useful tool for analyzing light refraction at an interface is the isoenergy map plotted in $k$-space. Given a dispersion relation $E(k_x,k_y)$, an isoenergy contour defines the curves over which the energy is constant. In this representation the gradient of the energy surface is the vector field of group velocities, which points normal to the isoenergy contour as illustrated by the colored arrows in \fig{fig:airlatt}(a). Using \eq{eq:energy_spectrum}, the group velocity $\vec{v}_g^\pm \equiv \nabla_{\vec{k}}E^\pm$ is expressed as,
\begin{equation}
			\vec{v}_g^\pm = [-1 \mp \frac{\Delta - K}{\sqrt{(\Delta - K)^2 + 4\beta^2}}]\nabla_{\vec{k}} K/2.
\label{eq:group_velocity}
\end{equation}

AANR of single photons can take place at the interface between free-space and a JCH lattice. Consider the band structure of a rotated lattice shown in \fig{fig:airlatt}(a) which is superimposed with the lattice and free-space isoenergy contours of matching energy. The contours associated with free-space are circles with radius equal to its energy (in natural units). By requiring that the surface parallel wave vector $k_y$ is conserved at the interface, the group velocities associated with these contours determine the refraction angle. In the illustration, an incident photon with wavevector $\vec{k}_1=(k_{1,x},k_{1,y})$ and velocity $\vec{v}_1$ will couple to an allowed mode of the lattice, and propagate with $\vec{k}_2=(k_{2,x},k_{2,y})$ and $\vec{v}_2$. The refraction angle is
\begin{equation}
	\theta_R = \arctan(\tan k_{1,y} \cot k_{2,x})~,
\label{eq:refraction_angle}
\end{equation}
where $k_{2,x}$ is given by \eq{eq:k2_condition}. Since the isoenergy contours of the lattice are \emph{convex}, we have negative refraction ($\theta_R < 0$), and since the lattice contour is larger than the air contour, this occurs for all incident angles.

To converge the light the isoenergy contour needs to be as circular as possible. This occurs at the energy band extrema, and so for a sharp focus it is preferable to work as close to these frequencies as possible.

Refraction is not invariant under lattice rotation. This is because there is a change in the air-lattice interface under rotation. From the conservation of the $k_y$ condition, Fig.~\ref{fig:airlatt}(b) shows that the unrotated lattice does not exhibit AANR. When AANR does occur, the propagating modes in free space can be brought into a focus to form an image on the other side of the lattice even with a planar lattice slab. Such a device therefore satisfies the first criterion of a superlens. In contrast, conventional lenses that rely on positive refraction must have curvatures to converge light. 

An exact numerical simulation to test the predicted negative refraction (Eqs.~(\ref{eq:group_velocity}) and (\ref{eq:refraction_angle})) for an air-lattice interface requires knowledge of, and is dependent on, the specific light-cavity coupling mechanism at the interface of a physical implementation. To demonstrate the underlying principles discussed above without recourse to a specific coupling mechanism, we conduct our numerical simulation at an interface between two JCH lattices, with no loss in generality.

\begin{figure}[htbp]
	\centering
	\includegraphics[width=0.8\columnwidth]{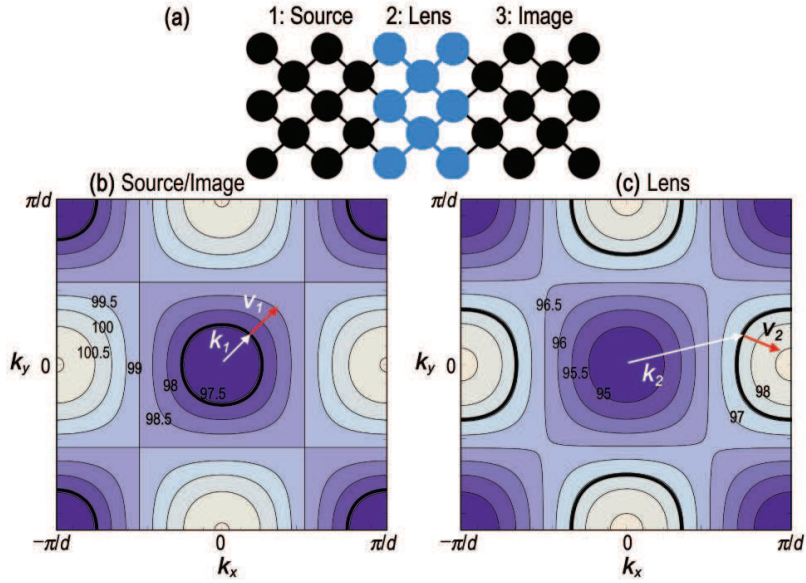}
	\caption{Energy band structure with lattice-lattice isoenergy contours. (a): Lattice configuration for numerical simulations. Band structures for (b) source and image, and (c) the lens. The regions are distinguished by their respective atomic detuning $\Delta$, and negative refraction is predicted at the lens interfaces. The parameters are $\Delta_1 = 0, \beta = 100\kappa, \Delta_2 = -5.27\kappa$.}
\label{fig:lattlatt}
\end{figure}

We use a segmented lattice as our platform (depicted in Fig.~\ref{fig:lattlatt}(a)), where our single photon source is initialized in the `source' region, and there is an identical `image' region which acts as the image plane. The sandwiched `lens' region  negatively refracts the excitation and brings it to focus in the image plane. The band structures for the source and lens region are shown in Figs.~\ref{fig:lattlatt}(b) and~\ref{fig:lattlatt}(c) respetively. For the chosen operating frequency, indicated by the bold isoenergy contours, the system exhibits AANR.

\begin{figure}[htbp]
	\centering
\includegraphics[width=0.8\columnwidth]{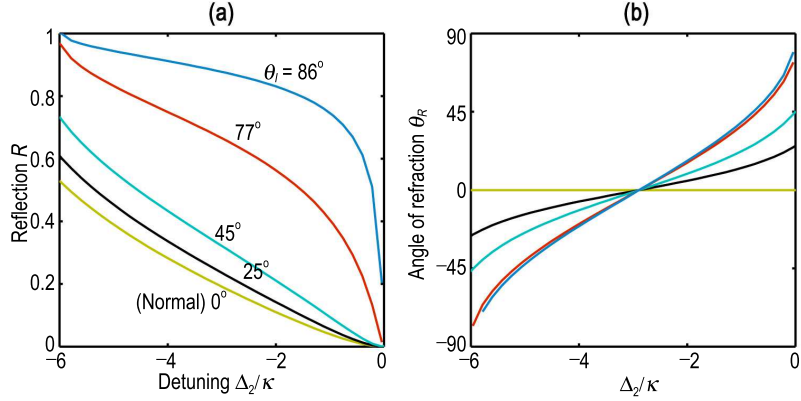}
\caption{Trade-off between reflection and refraction angle as a function of atomic detuning. (a) Reflection coefficient, $R$, and (b) refraction angle, $\theta_R$, as a function of lens atomic transition energy, $\Delta_2$ for different incident angles, $\theta_I$ (note that same colored curves correspond to same incident angle). Large negative refraction is accompanied by large reflection. The model parameters follow Fig.~3.}
	\label{fig:RTheta}
\end{figure}

Due to the dielectric mismatch of the interface, there must be a finite probability of reflection. At the interface, the discrete scattering eigenequation can be used to derive the reflection coefficient (see \textsc{appendix}),
\begin{equation}
		R = \frac{1 - \cos(k_{1,x}-k_{2,x}) } {1 - \cos(k_{1,x}+k_{2,x}) }
\label{eq:reflection}
\end{equation}
and from conservation of energy the transmission coefficient $T \equiv 1-R$. Comparing \eq{eq:refraction_angle} and \eq{eq:reflection}, there is a trade-off between refraction and reflection, i.e, large negative refraction is accompanied by large reflection. This trade-off is illustrated in a comparison plot (~\fig{fig:RTheta}) of the refraction angle and reflection coefficient for different incident angles and varying detuning.

\begin{figure}[tb!]
	\centering
	\includegraphics[width=1\columnwidth]{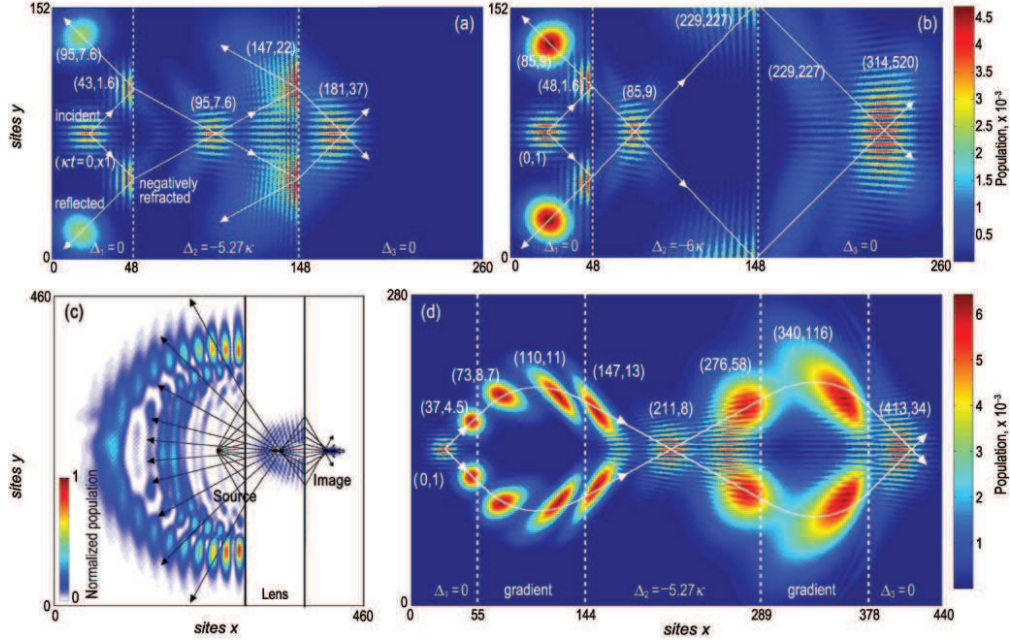}
	\caption{Negative refraction of Gaussian polaritonic pulses. The pulse is initialized with a coherent superposition of two momenta $\vec{k}_1 = (\pi/4,\pm\pi/4)$. We superimposed different time instances $t$ of system evolution with each instance labeled with $(\kappa t, M)$ where the population is multiplied with $M$ for a clearer presentation. (a): The atomic transition energy in the lens is $\Delta_2 = -5.27\kappa$. (b): $\Delta_2 = -6\kappa$. Predicted trajectories are indicated by the arrows. The polariton follows the predicted trajectories for incidence, reflection and refraction. Changing $\epsilon_2$ changes the focal point. (c): A snapshot at time, $t = 300 / \kappa$, of the imaging of a point-like source by negative refraction (Media 1). (d): The lattice implements a gradient-index lens by employing adiabatic variations in $\Delta_2(x)$, reducing reflection (Media 2).  The other parameters follow Fig.~\ref{fig:lattlatt}.}
	\label{fig:simulation}
\end{figure}

The propagation of the field in the lattice is governed by the Schr\"{o}dinger equation $|\psi(t)\rangle = e^{i \mathcal{H}t} |\psi(0)\rangle$. We consider the case when the source is initialized in an equal superposition of atomic and photonic modes. It is instructive to use a directional pulse by specifying an initial  state with a normalized Gaussian momentum distribution around $\vec{k}_1 = (\pi/4,\pm \pi/4)$ so that it is incident on the lens at $\pm45^\circ$, as shown in Fig.~\ref{fig:simulation}(a). The superposition of the two $\vec{k}$ modes manifests in a coherent interference pattern in the $y$-direction.

The lens atomic detuning is set to $-5.27\kappa$, which by \eq{eq:refraction_angle}, predicts a refraction angle of $\theta_R = -25^\circ$. Superimposing different time instances, the incident, reflected and refracted pulses in \fig{fig:simulation}(a) follow the predicted refraction angle and the trajectory predicted by \eq{eq:group_velocity}, to converge at a location on the image plane. The reflection and transmission coefficients are also found to be in good agreement with \eq{eq:reflection_coefficient_effective}. The incident and reflected polariton $(\pm \pi/4,\pi/4)$ coherently interfere near the interface to give an interference pattern along the $x$-direction. Note that there is considerable reflection, so that the population density has been multiplied by a factor $M$ in \fig{fig:simulation}(a) and \fig{fig:simulation}(b) for clearer representation.

An important property of our system, distinct from the existing PhC superlens implementations, is the ability to tune the atomic transition energy, $\epsilon$, after fabrication. Such manipulations can be achieved dynamically by, for example, a controlled external electric field via Stark shift.  This control allows one to tailor the dispersion relation, and hence the light guiding properties and focal point of the lens. The effect of changing $\epsilon$ is demonstrated in Fig.~\ref{fig:simulation}(b), where decreasing $\Delta_2$ by $0.73\kappa$ shifts the focus 56 sites to the right.

To show how a point-like source would converge with AANR onto the image plane, we specify a source with an initial superposition of Gaussian momentum distribution summed over $\vec{k}$ in \fig{fig:simulation}(c). Taking a snapshot of the propagation at time, $t=300/\kappa$, it shows that as the point source propagates into the planar lens, all components are negatively refracted, so that an image of the point source is successfully formed on the image plane. As with \fig{fig:simulation}(a) and \fig{fig:simulation}(b) there is considerable reflection at the lens' interfaces.

In our examples so far, the lenses are homogenous and there is an abrupt change at the interface. As a result, there is considerable reflection such that the total transmission through the lens is less than 25\%. As expressed by \eq{eq:reflection}, reflection increases with the greater the change in $k_{x}$. This can be minimized if we provide an adiabatic spatial change of the atomic transition energy within the lens, in effect producing a gradient-index (GRIN) structure. In \fig{fig:simulation}(d) the detuning distribution follows the form,
\small
\begin{equation}
\Delta(x')=\begin{cases}
	(\Delta_2 - \Delta_1) \sin^2(\frac{\pi x'}{2 w})+\Delta_1& \text{if $0 < x' \leq w$},\\
	\Delta_2& \text{if $w < x' \leq W-w$},\\
	(\Delta_2 - \Delta_1) \cos^2(\frac{\pi (x'+W-w)}{2 w})+\Delta_1& \text{if $W-w < x' \leq W$}.
\end{cases}
\end{equation}
\normalsize
where $x'$ is the number of sites from the interface, $w$ is the width of the GRIN region and $W$ is the total width of the lens. By fine tuning the GRIN region, the level of reflection can be made arbitrarily small, although the physical trade-off is a larger lens. The removal of reflection losses is an important development, demonstrating fine control of propagation possible in our system.

\section{Evanescent wave enhancement}
The ability of lenses to resolve images is limited by the wavelength of the light source because the high-spatial-frequency modes that describe the subwavelength features are non-propagating and do not reach the image plane. To see this, the dispersion relation in free space, $k_x = \sqrt{\omega^2 - k_y^2}$, implies that the modes with $k_y > \omega$ exponentially decay away from the source along the $x$-axis. Existing superlens proposals overcome this diffraction limit by amplifying the evanescent wave (EW) components.

PhC-based evanescent wave enhancement (EWE) devices can be regarded as a type of resonator. At resonance, the transmission of the evanescent components is divergent. The total transmission across a lens of width $W$, derived from taking the summation of the multiple scattering events at the left and right interfaces is,
\begin{equation}
	\mathcal{T} = \frac{T_{12}T_{23}R_{23}}{\exp(-2ik_{2,x}W)-R_{23}^2}~,
\label{eq:transmission}
\end{equation}
where $T_{ij}$ ($R_{ij}$) is the transmission (reflection) amplitude at the interface between region $i$ and $j$. At the resonance condition $\exp(-2ik_{2,x}W)-R_{23}^2 = 0$, transmission is divergent. The resonant condition is just the condition for total internal reflection where the accumulated phase shift in a round trip is a multiple of $2\pi$.

The resonant \emph{bound modes}~\cite{luo03a} allow a build-up of these bound states to produce an amplified evanescent tail on the image side of the lens. These resonant modes are discrete, therefore they will not amplify all evanescent modes. NIM-based superlenses do not have this limitation, but because their fabrication is based on conducting elements they suffer the problem of loss.

Since JCH-based superlenses can dynamically shift the resonant points, they can overcome the limitation of PhC-based superlenses and amplify a contiguous range of evanescent modes (but not simultaneously).

Resonant bound modes can either exist along the interface (surface bound modes) or in the bulk (bulk bound modes). As both mechanisms follow the same underlying resonance principle, we demonstrate only the latter. 

\begin{figure}[tb!]
	\centering
	\includegraphics[width=0.7\columnwidth]{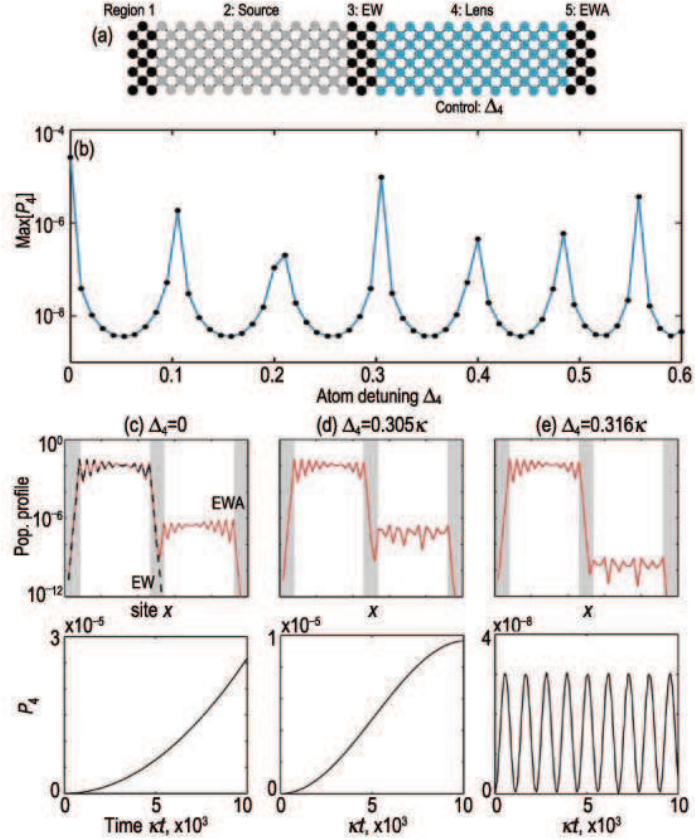}
\caption{EWE by tuning atomic transition energy. (a): Lattice schematics for demonstrating EWE. The evanescent wave (EW) is initialized in the region 3 by preparing an eigenstate of energy $\mathcal{E}-\omega \sim 100\kappa$ in the region 1--3. The lens serves to amplify the field via resonant coupling. (b): Population buildup in the lens ($P_4$) occurs at the quasi-resonances (peaks). (c)--(e): Population profile taken along the $x$-axis and time evolution of population $P_4$ for selected $\Delta_4$ values. Dashed blue line shows the rapid drop-off of the EW without the lens, such that the incident population or probability would be $10^{-12}$. The lens provides the enhancement on resonance seen in (c) and (d). The parameters are $\Delta_{1,3,5} = 0$, $\Delta_{2} = 0$ and $\beta= 100\kappa $. }
	\label{fig:EWA_eps}
\end{figure}

A quantum equivalent of an EW is the evanescent tail of a stationary state. We prepare our system such that the detuning in regions 1 and 3, $\Delta_1$ and $\Delta_3$ respectively, are sufficiently different from the detuning in region 2, $\Delta_2$ (see \fig{fig:EWA_eps}(a)), so that we can setup an eigenstate where evanescent tails exists in region 1 and 3. This setup is analogous to that of a square well. Region 4 is the lens, and region 5 our image plane.

The resonant condition for EWE can be achieved by tuning the atomic transition energy of the atoms in the lens region or the width of the lens. 
We will only consider the former here, as it is more appropriate to the idea of a reconfigurable device. We solve for the time evolution of the system and observe in Fig.~\ref{fig:EWA_eps}(b) that significant coupling to the lens occurs at distinct values of the atomic detuning, $\Delta_4$, where the resonance condition is met. 

When the lens is exactly on resonance, as demonstrated in \fig{fig:EWA_eps}(c) where $\Delta_2 = \Delta_4 = 0$, the population exchange occurs between the source and the lens is that of coupled homogenous resonators, i.e., $P_4(t) = \sin^2(\Omega t)$, where $\Omega$ is the characteristic mutual coupling. Following the increase in $P_4$, the EW incident on the lens is transmitted amplified by 6 orders of magnitude after a time $t = 10^4 \kappa^{-1}$.

When the lens is only quasi-resonant (Fig.~\ref{fig:EWA_eps}(d)), the population exchange between the coupled resonators (source and lens) can be approximated by,
\begin{equation}
	P_4(t) = \frac{2\Omega^2}{\eta^2 + 4\Omega^2}\Big[1 - \cos(\sqrt{\eta^2 + 4\Omega^2}t) \Big]~,
\end{equation}
where $\eta$ is the difference in the eigenenergy of the source and lens. By fitting $P_4(t)$ to numerical results, we find that for $\Delta_4 = 0.305\kappa$, $\eta \sim 10^{-3}\kappa$. This is in good agreement with the minimum energy difference between source and lens obtained by solving the lens Hamiltonian $\mathcal{H}$ directly. 

\fig{fig:EWA_eps}(c) shows that at $t = 10^3 \kappa^{-1}$ the incident EW is amplified by a third of the exact resonance case. Thus, although exact resonance is an optimal condition for EWE it is not a necessary condition for enhancement. When the lens is tuned away from resonance, the degree of EWE can quickly diminish as seen in Fig~\ref{fig:EWA_eps}(e).

The diffraction limit restricts the resolution of conventional lenses to the operating wavelength, $\Lambda_0$. Our lens' resolution, $\delta = 2\pi / k_{\rm{max}}$, is determined by the maximum $k$ that still satisfies the resonant condition. However we would also like to resolve all the $k$-components leading up to $k_{\rm{max}}$. This implies minimizing the bulk energy band spectrum so that the deviation from the resonant energy is always small. The drawback of this is a reduction in sharpness of focus. A better solution is to introduce surface mode resonance. This can be achieved by having a different $\epsilon$ at the lens surfaces from that of the bulk. As shown in \fig{fig:EWA_bands}, the \emph{flatter} surface mode band (see appendix) provides the necessary minimal deviation from resonance to maximize $k_{\rm{max}}$, leaving the bulk mode to provide the AANR and focal sharpness.

\begin{figure}[tb!]
	\centering
	\includegraphics[width=0.8\columnwidth]{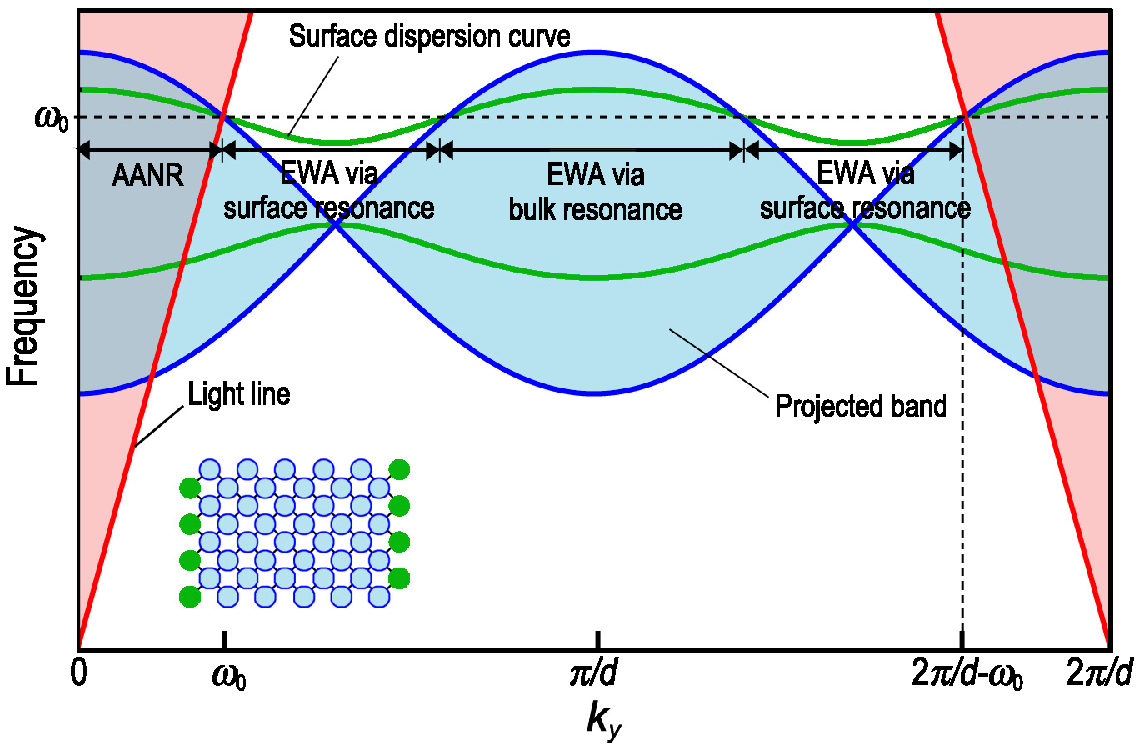}
	\caption{Surface and bulk energy band structures for evanescent wave enhancement. There are regions in the bulk energy spectrum which deviate from the frequency, $\omega_0$. In these gaps, EWE via bound bulk mode resonance can not occur. By having surface mode resonance, these gaps can be filled. The thin surface energy spectrum around the operating energy, $\omega_0$, maximizes the range of $k$ for which EWE can occur, whilst the broad bulk energy spectrum	provides AANR and focal sharpness. Surface modes can be achieved by independently tuning the lens surfaces from the bulk (as depicted in the inset).}
	\label{fig:EWA_bands}
\end{figure}

\fig{fig:EWA_bands} shows that $k_{\rm{max}} = 2\pi / d - \omega_0$, because beyond this the evanescent modes fold back into the light cone and the associated bound modes become \emph{leaky} states~\cite{luo03a}. Thus the maximum resolution of our lens is,
\begin{equation}
	\delta = \frac{d}{1 - d / \Lambda_0}~,
\label{eq:resolution_def}
\end{equation}
so that for sufficiently small inter-cavity spacing, $d < \Lambda_0 /2$, the resolution exceeds that of conventional lenses.

Inter-cavity spacing is however limited by the size of the cavity. This means that to beat the diffraction limit, one at the very least needs the cavity size to be less than $\Lambda_0$. Typically cavity resonant wavelength is twice the cavity size, so the subwavelength resolution condition becomes, $\omega > \omega_0$. The non-linear interaction introduced by the cavity atom allows, beyond that which is available through just inter-cavity hopping, the cavity resonance frequency to be greater than the operating frequency.

Using \eq{eq:energy_spectrum} where $E^\pm = \omega_0$, the relative resolution of our lens can be approximated by,
\begin{equation}
	\frac{\delta_0}{\delta^\pm} \approx \frac{\omega \mp \beta}{\omega \pm \beta}~,
\label{eq:relative_resolution_short}
\end{equation}
where we have assumed small detuning and that the minimum possible spacing between sites is half the resonant wavelength. Equation~(\ref{eq:relative_resolution_short}) gives the factor by which our lens beats the diffraction limit. Since $\delta_0 / \delta^+ < 1$, only the  resolution from the negative energy branch, $\delta^-$, can better the diffraction limit. Conventionally $\beta/\omega$, restricted by the so-called fine structure constant limit, is of the order 0.01 (although larger values are possible for unconventional coupling mechanisms~\cite{Devoret07}), so the improvement over the diffraction limit is typically small.

\section{Experimental feasibility and outlook}

The lead time from theoretical inception to experimental realization can be long. NIMs were originally proposed in 1968~\cite{veselago68}, but it was not physically realised until 2000~\cite{smith00}. The case for the experimental feasibility of CAMs is fortunately alot clearer. Optical microcavities have been created with whispering gallery modes (WGM)~\cite{LefevreSeguin97}, fabry-perot microcavities~\cite{hood01,kohnen09}, photonic bandgap (PBG) defects~\cite{Reese01}, and in slot waveguides~\cite{sun07}. The coupling of PBG nanocavities into a one-dimensional array has also been achieved~\cite{notomi08}. Recently the coupling of Nitrogen-vacancy (NV) centers to PBG cavities~\cite{barth09} and WGM  microdisks~\cite{barclay09} have been experimentally verified. With the astonishing advancement in microcavities and arrays of microcavities, and their coupling to multi-level quantum systems, it is feasible that the next advancement in experimental development would be that of arrays of microcavities coupled to multi-level quantum systems, in other words, CAMs. The dynamic control of Stark tunable solid-state emitters have been been demonstrated with diamond colour centres~\cite{tamarat06}, quantum dots~\cite{empedocles97} single molecules~\cite{brunel99}, and single-ion dopants~\cite{alexander06}.

The development of microcavities can be characterized by their size and quality factor, $Q$. WGM microcavities have experimentally produced $Q \sim 10^9$ in microspheres~\cite{Gorodetsky96} and $10^8$ in microtoroids~\cite{Armani03, Spillane05}. However their sizes are relatively large: $10^3$ $\mu\rm{m}^3$ and $180$ $\mu\rm{m}^3$ respectively. PBG microcavities have achieved $Q \sim 10^7$ with cavity mode volume, $V \sim (\lambda/2)^3$, where $\lambda$ is the operating wavelength~\cite{Noda07}. NV couplings in PBG microcavities have been calculated as $\beta \sim 10^{10}$ Hz~\cite{Hiscocks09}. Assuming that photon hopping limits $Q$, we can approximate the inter-cavity tunneling frequency as $\kappa = \omega / Q$. For the superlensing properties presented in this work, $\beta = 100\kappa$, requiring in the visible light regime $Q \sim 10^7$, which is at current experimental limits. In PBG arrays with over 100 microcavities however, only $Q \sim 10^6$ as yet been experimentally verified~\cite{notomi08}. Coupling subwavelength-sized slot-waveguide cavities in 2D lattice has also been discussed~\cite{su10}.

\begin{figure}[tb!]
	\centering
	\includegraphics[width=0.8\columnwidth]{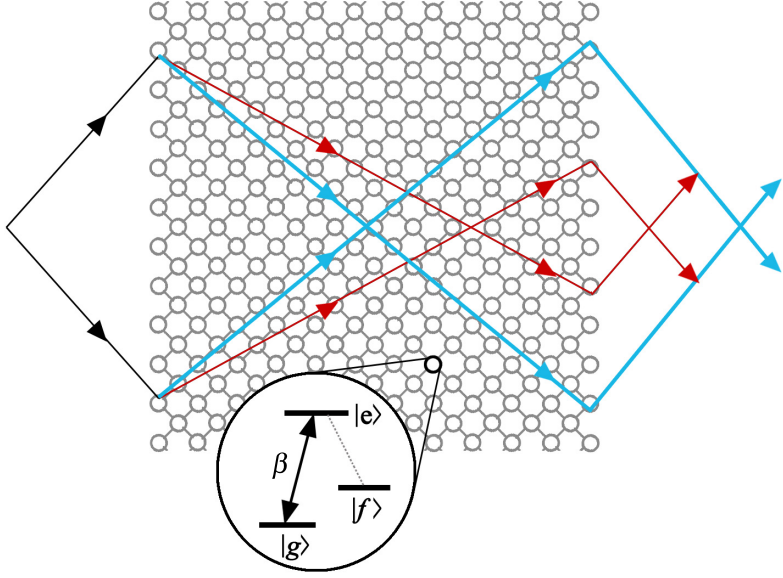}
	\caption{Quantum superposition of material properties with cavity $\Lambda$-system. Each cavity is a $\Lambda$-system (e.g. quantum dots or NV centers in diamond) consisting of two atomic ground states and one excited state, $|e\rangle$. One of the ground states, $|g\rangle$, is coupled to the cavity mode via coupling parameter $\beta$ and the other ground state, $|f\rangle$, is only weakly coupled. When some of the atoms are in a GHZ-like state, the metamaterial will exhibit a superposition of dispersion relations. The superlens then would exhibit a superposition of two focus points, represented by the blue and red arrows.}
	\label{fig:lambda_system}
\end{figure}

Atom-optical microcavities have demonstrated many quantum effects ranging from quantum collapse and revival to eletromagnetically induced transparency. It is upon these quantum effects that CAMs' true potential lies. In particular, a transmission line formed by coupling superconducting charge qubits prepared in a coherent superposition of quantum states has been studied to show an oscillating band gap and transmittance~\cite{rakhmanov08}. First experimental steps to realize such materials have been reported ~\cite{abdumalikov10,astafiev10a,astafiev10b}. Along similar lines for future investigation, coupling of a three level $\Lambda$-system (e.g. using quantum dots~\cite{xu08} or diamond defects~\cite{santori06}), to the cavity modes could yield a superposition of dispersion relations and thus create a metamaterials with a quantum superposition of material properties. One of the ground states ($|g\rangle$) in the individual cavity $\Lambda$-system would be strongly coupled to the cavity modes while the other ($|f\rangle$) is not~(\fig{fig:lambda_system}). The atom-photon coupling offers intriguing potential for entirely new quantum devices.  For example, if some of the atoms are prepared in a Greenberger–-Horne-–Zeilinger--like (GHZ-like) state $|gg...g\rangle + |ff...f\rangle$, then the system will exhibit a superposition of dispersion relations.  This in turn implies a superlens with two focal points in quantum superposition.

\section{Conclusion}
In summary, we have combined the previously unrelated fields of quantum mechanics and metamaterials, by proposing cavity arrays as a new class of dynamic metamaterial. Operating at the quantum level, it opens up new possibilities for quantum optical devices. By applying traditional metamaterial techniques we showed that the CAM can exhibit the features of a superlens. In a more general sense, this work lays down the framework for local manipulation of photons, the quantum superposition of metamaterial properties, the preservation and interaction with entangled fields, and other non-local effects, in cavity array metamaterials, creating a new area of investigation in quantum transformation optical phenomena. This invites quantum technology into the realm of metamaterials.

\appendix	
\section*{Appendix}
\noindent 1. The method described here for the derivation of the band structure solution in \eq{eq:energy_spectrum} follows Ref.~\cite{quach09}. In the presence of intercavity coupling, the onsite energies $E_r$ are no longer the polaritonic energies $E_n^\pm$ and in general, satisfy the relation,
\begin{equation}
	\sum_{s}{H_{rs} \vert \phi,s \rangle = E_r \vert \phi,r \rangle},
\label{eq:energy_general}
\end{equation}
where  $H_{rs}$ is the Hamiltonian that relates site $r$ to site $s$. Employing Bloch's theorem for periodic structures,
\begin{equation}
	\vert \phi,s \rangle = \vert \phi,r \rangle \exp[i \vec{k}\cdot(\vec{d}_s-\vec{d}_r)],
\label{eq:discrete_bloch}
\end{equation} 
where $\vec{d}_r$ is the displacement to site $r$, and $\vec{k} \equiv (k_x,k_y)$ is the wavevector associated with the crystal momentum, \eq{eq:energy_general} becomes an energy eigenequation whose eigenvalues are the energy band structure or the dispersion relation of the medium.

\noindent 2. For an incident field of energy $E_1$ and wavevector $\vec{k}_1=(k_{1,x},k_{1,y})$ with transmitted field of $E_2$ and $\vec{k}_2=(k_{2,x},k_{2,y})$ at the interface, energy conservation $(E_1 = E_2)$ and phase matching ($k_{1,y}=k_{2,y}=k_{y}$) requires that $k_{2,x}$ satisfies the condition,
\begin{equation}
		K = \omega_2 - E_1 + \beta_2^2 / (E_1 - \epsilon_2)~,
\label{eq:k2_condition}
\end{equation}
where $K \equiv 4\kappa\cos(k_{2,x} d)\cos(k_{1,y} d)$ for the rotated lattice.

\noindent 3. For the derivation of the reflection coefficient (\eq{eq:reflection}), the state vector can be expanded in the bare atom-photon basis,
\begin{equation}
	|\psi\rangle = \sum_r (c_r|g,1\rangle_r + d_r|e,0\rangle_r) \bigotimes_{s\neq r} |g,0\rangle_s
\end{equation}
where $|g,n\rangle_r$ and $|e,n\rangle_r$ denote the ground and excited state respectively, with $n$ photonic excitations at site $r$. Given the symmetry, we consider a 5-site unit cell in X-configuration that is translational-invariant along the $y$-direction. Using the standard eigenenergy equation $\mathcal{H}|\psi\rangle = E|\psi\rangle$, one arrives at a discrete scattering equation for each region of the lattice and the interfaces. In particular in region $j$ with associated parameters $(\omega_j,\epsilon_j,\beta_j,\kappa_j)$, we have
\begin{equation}
	\kappa_j(c_{p,q-1} + c_{p,q+1}+ c_{p+1,q} + c_{p-1,q}) = \Big( \omega_j + \frac{\beta_j^2}{E-\epsilon_j} - E\Big) c_{p,q},
	\label{eq:scattereq}
\end{equation}
where a given site at coordinate $(p,q)$ is surrounded by four nearest-neighboring sites at coordinates $(p\pm 1,q\pm 1)$ and the conservation of energy requires $E_j = E$. At the interface centered at the origin (0,0), 
\begin{equation}
	\kappa_1(c_{0,-1} + c_{1,0}) + \kappa_2(c_{0,1}+c_{-1,0}) = \Big( \omega_2 + \frac{\beta_2^2}{E_2-\epsilon_2} - E\Big) c_{0,0}~.
	\label{eq:interfeq}
\end{equation}
We make the typical assumption that region 1 consists of an incident and a reflected wave component,
\begin{equation}
	c_{p,q} = e^{i k_{1,p} p}e^{i k_{1,q} q} + r e^{-i k_{1,p} p}e^{-i k_{1,q} q}~,
\end{equation}
where $k_{j,p} = k_{j,x}-k_{y}$ and $k_{j,q} = k_{j,x}+k_{y}$. In region 2, the transmitted wave has the form,
\begin{equation}
	c_{p,q} = t e^{i k_{2,p} p}e^{i k_{2,q} q}~,
\end{equation}
where $r$ and $t$ are used here to denote reflection and transmission amplitudes respectively. Substituting these solutions in to the interface equation (\eq{eq:interfeq}) and applying continuity condition $t = 1+r$, we arrive at the reflection coefficient $R \equiv |r|^2$,
\begin{equation}
		R = \frac{\kappa_1^2 + \kappa_2^2 - 2 \kappa_1\kappa_2\cos(k_{1,x}-k_{2,x}) } {\kappa_1^2 + \kappa_2^2 - 2 \kappa_1\kappa_2\cos(k_{1,x}+k_{2,x}) }.
\end{equation}
Assuming uniform coupling $\kappa_j = \kappa$, we retrieve the required expression. 
Finally since a polaritonic pulse has a momentum distribution $G(\vec{k})$, we define an effective reflection coefficient,
\begin{equation}
	R_{\rm eff} = \int_{-\pi}^\pi\int_{-\pi}^\pi G(\vec{k}) R(\vec{k})d\vec{k}.
\label{eq:reflection_coefficient_effective}
\end{equation}

\noindent 4. To calculate the surface mode energy band, we need to take two adjacent sites along the surface as the primitive cell. After applying Bloch's theorem the Hamiltonian is,
\begin{equation}
	\mathcal{H}=\begin{pmatrix}
	\omega&	\beta&	-\kappa[1+\exp(i k_y \sqrt{2}d)]&	0\\
	\beta&	\epsilon_s&	0&	0\\
	-\kappa[1+\exp(-i k_y \sqrt{2}d)]&	0&	\omega&	\beta\\
	0&	0&	\beta&	\epsilon_b
	\end{pmatrix}~,
\label{eq:}
\end{equation}
where $\epsilon_s$ and $\epsilon_b$ are the surface and bulk atomic transition energies respectively. We solve for the eigenvalues to get the surface mode energy bands.

\section*{Acknowledgements}
This work was supported by the Australian Research Council (ARC) under the Centre of Excellence scheme.  L.C.L.H.  and A.D.G. acknowledge the financial support of the ARC under Projects DP0770715 and DP0880466, respectively.

\end{document}